\documentclass[twocolumn]{NobArticle}
\runninghead{Telescope proposal writing}

\title{Valuation system for telescope observation time proposals for the Gemini Observatory}

\author{
C.G. D\'iaz\textsuperscript{1,2},
R. Petrucci\textsuperscript{1,2},
L.V. Ferrero\textsuperscript{2} \&
E. Jofré\textsuperscript{1,2} 
}
\date{
    \textsuperscript{\textbf{1}}
    Consejo Nacional de Investigaciones Cient\'ificas y T\'ecnicas (CONICET), Godoy Cruz 2290, CABA, CPC 1425FQB, Argentina\\ 
    \textsuperscript{\textbf{2}}
    Universidad Nacional de C\'ordoba, Observatorio Astron\'omico de C\'ordoba, Laprida 854, C\'ordoba X5000BGR, Argentina 
    }

\renewcommand{\maketitlehookd}{%
\begin{abstract}
    \noindent The preparation of a telescope observation time proposal is a recurring activity in observational astronomy. It is a necessary investment of time and effort to obtain data to advance a research topic. Therefore, the success of an observation proposal is a condition for progress in observational research. 
    This guide was created to offer a straightforward, practical, and comprehensive resource for applicants who are preparing a Gemini Observatory proposal. It reviews the fundamentals of an observation proposal, including its content and evaluation criteria, to help applicants organize their submissions effectively and improve specific aspects of their presentations.
    This manuscript is inspired by the recommendations of the User Advisory Council established in the documents \textit{``Criterios de evaluación de propuestas por parte del NTAC'' }(National Time Allocation Committee), \textit{``Consideraciones básicas para la presentación de propuestas Gemini''}
    and in experiences of members of the NTAC.    
    \medskip
    
    \small{\textbf{Keywords:} Scientific Writing, Telescope Proposal, Gemini Observatory.}
\end{abstract}
}

\begin{document}

\small
\maketitle

\section{Introduction}
An observation proposal is a brief presentation that describes an experiment whose design must follow certain logic.
The brevity of the document should not be a limitation, it simply implies certain degree of optimization in the presentation of relevant information, such as: avoiding repeating information, respecting the content of each section, using only those figures that contribute to the interpretation of the proposal, using tables correctly, etc.
Furthermore, since it is a scientific experiment, it needs minimum content elements such as theoretical framework, hypothesis, objectives, and methodology.
Since telescope time is extremely valuable, the observing proposal requires serious planning in order to find a strategy that is efficient and effective in contributing to the scientific objective.
Therefore, there is a long way to go from a very good idea to a good, clearly justified observation proposal.

In particular, for the Gemini Observatory, an observing proposal should have three main sections\footnote{See section
\href{https://www.gemini.edu/observing/phase-i/pit/pit-description\#PDF}{\it ``The PDF Attachment''}
from the Phase I Tool (PIT) on the Gemini web page \citep{pdfattach}.}:
Scientific Justification (SJ, \ref{sec_jc}), Experimental Design (ED, \ref{sec_de}), and Technical Description (TD, \ref{sec_dt}).
The Scientific Justification serves as a theoretical framework and background review.
The problem that is being addressed must be clearly stated
and the reader must be able to tell
how the scientific objective of the proposed project will contribute to the scientific case.
The Experimental Design is the section that contains the methodology of the experiment, and must describe the strategy, sample, and tools for analysis once the data is acquired. This section is important since it allows the investigators to demonstrate that they have not only knowledge of the science topic, but also of how the data will be analyzed.
Finally, the Technical Description contains all the information necessary for the acquisition of the data, and as such, it must provide all the relevant details to make the observations requested in the proposal.
In particular, this section is so important that the Gemini National Office of each partner provides a specialized review of it (for each proposal), carried out by peers familiar with the operation and characteristics of each requested instrument.
In general, it is considered that the three sections mentioned have the same weight in the evaluation of a proposal. However, a serious technical problem can disqualify a very strong, high-impact science case.
Furthermore, the quality of the experimental design can only be appreciated if the scientific justification is complete, easy to follow, and relevant. Therefore, there are certain parameters that must be considered in preparing a proposal to ensure good performance in the time allocation process.

Generally, during the evaluation process each section is scored on a point scale and the total points of the proposal will give it its place in the final ranking \citep{criteriosNTAC}.
Therefore, proposals must be documents that are easy to read, easy to score, and understandable by colleagues from other areas without prior knowledge of the topic \citep{consideracionesPI}.
To do this, it is important to optimize the presentation of the information so it is easily accessible.
This document presents a review of the content of each section of a Gemini Observatory observation proposal and the aspects to be considered in each of them.
In addition, for each section, a simple and justified valuation system is proposed according to content and quality. This rubric can be used to identify ways to move towards an orderly, complete, and experimentally convincing presentation.

\section{Main sections of an observation time proposal}\label{s:contenido}

The characteristics to be assessed in the three main sections of a Gemini proposal are presented below. Detailed criteria can be found in the rubric (Section \ref{s:rubric}) which provides a quick and easy way to check the completeness of each part of the presentation.

\subsection{Scientific Justification} \label{sec_jc}
The Scientific Justification (SJ) section must include the following:
\begin{enumerate}
\item {\bf Scientific context: \label{context_cient}} Presentation of the research topic.
All the necessary information
(results of previous research, references, figures, etc.) must be provided for the correct interpretation of the other aspects of this section. For example, the relevance of the problem or the connection to the overall objective cannot be assessed without a clear context.
\item {\bf Problem relevance: \label{relevance}} A measure of the importance of the problem addressed within the scientific context provided.
\item {\bf General Objective: \label{obj_generla}} The overall contribution of the proposal to the scientific case, the sum of the specific objectives.
\item {\bf Specific or Immediate Objectives: \label{obj_espec}} Individual contributions obtainable from the analysis of the data and which constitute the basis of the general objective.
\item {\bf Impact: \label{impact}} Measure of the scope of the objective(s) within the context provided.
\end{enumerate}
\subsection{Experimental Design} \label{sec_de}

After having presented the scientific case and the objectives, it is time to describe the particularities of the experiment.
The expected content of the Experimental Design (ED) section is as follows:

\begin{enumerate}
\item {\bf Sample: \label{sample}} The objects to be observed are presented, justifying their selection in view of the objectives set out in the previous section.

\item {\bf Strategy: \label{strategy}} A description of the measurements that will be made with the data, and how it is planned to extract the information necessary for the analysis to address the specific objectives.

\item {\bf Analysis tools: \label{tools}} The tools involved in the analysis of the measurements, in addition to the reduction of the data.
For example, light profile decomposition, spectral lines fitting, light curve fitting, statistical analyses, comparison with theoretical models, etc.

\item {\bf Contribution to the general objective: \label{contribution}} 
The strategy presented must clearly relate to the general objective and justify its feasibility.\end{enumerate}
\subsection{Technical Description} \label{sec_dt}

The Technical Description (TD) section must encompass all the information needed to assess the feasibility of the ED. Therefore, it is crucial that this section is thorough, accurate, and well-supported. 
The applicants should read thoroughly the instrument technical specifications and the corresponding observing strategies in case additional information not covered by this document is required. In general, the expected content includes:
\begin{enumerate}
\item {\bf Instrument configuration: \label{configuration}} The complete configuration of the selected instrument must be reported to ensure consistency with the previous sections. This information varies based on the instrument and type of observation but typically includes the selected filters, gratings, slit size, central wavelength, read mode, image binning type, field of view, etc.

\item {\bf Angular and/or spectral resolution: \label{resolution}} Justification must be provided to demonstrate that the expected angular and/or spectral resolution for the requested conditions and instrument configuration is sufficient for the scientific objectives.

\item {\bf Observation conditions: \label{condition}} The conditions necessary to execute the proposed observation program are indicated in terms of image quality (IQ), cloud cover (CC), sky brightness (SB), and water vapor (WV). The requested observation conditions must be justified within the following percentiles: 20\%, 50\%, 70\%, 80\%, 85\%, and 100\% or ``{\it Any}''.
If necessary, air mass boundaries can also be included.
The associated factors are usually the necessary angular resolution, the precision in flux calibration, the intrinsic brightness of the object, etc.

\item {\bf \emph{Offsets}, \emph{nodding} or \emph{dithering}: \label{offsets}} The choice between offset, nodding, and dithering depends on the instrument, observation type (imaging or spectroscopy), wavelength, and object type. Therefore, it is important to read the instrument documentation and observer guides to identify the best strategy. For near-infrared (NIR) and infrared (IR) observations, proper sky sampling is essential for accurate sky subtraction. This means that point source spectroscopy requires nodding along the slit, typically in patterns like ABBA, while imaging necessitates a dither pattern, either random or a grid of small offsets. However, for extended objects, neither nodding along the slit nor small dithers are feasible, requiring a full offset to the sky within minutes of the science observation for both spectroscopy and imaging. In optical observations of extended objects, a full offset to a sky position might be needed, though the timing is less stringent compared to NIR/IR observations, allowing for sky sampling after the science observation is completed. Additionally, depending on the instrument, dither patterns for imaging and nodding along the slit for spectroscopy are often recommended to mitigate the impact of bad pixels or hot columns in the detector. Finally, in multi-CCD cameras, imaging of extended objects may require offsets to cover gaps between CCDs, and spectroscopy may need wavelength offsets to avoid gaps in wavelength coverage.

\item {\bf Signal to Noise Ratio (S/N): \label{s/n}}
It must be demonstrated that the required S/N is achievable in the requested exposure time using the Integration Time Calculator (ITC) tool of the corresponding instrument and attaching the result to the proposal. If the instrument does not have an ITC, a detailed explanation of how the exposure time was calculated to obtain the desired S/N must be provided, which will be reviewed by the Technical Office.

\item {\bf Total and minimum times: \label{times}} 
The time necessary to execute all science observations and the necessary \textbf{nighttime} ``baseline'' calibrations must be indicated as TOTAL TIME (does not include the baseline calibrations that are obtained during the day or at dawn/sunset).
The time necessary to achieve at least one scientific objective is the MINIMUM TIME. Its justification is closely linked to scientific objectives.
Both total and minimum time must include the \textit{overheads} times and must be consistent with the ITC examples, if applicable. It is useful and highly recommended to confirm these times with the \textit{``Observing Tool''} (OT) software, which is the tool for Phase 2 of the Gemini programs.
It is essential to keep in mind that the technical reviewer must be able to reproduce the calculation of the requested times (total and minimum) based on the information provided in the proposal.
Then, the values used for the \textit{offset}, read/write time, filter change time, among others, must be specified.
With respect to the minimum time, it is necessary to specify for which object or sub-sample of objects it was computed.

\item {\bf Calibrations: \label{calibrations}}
All calibrations necessary for the scientific case that were not included in the baseline calibrations provided by the observatory must be mentioned. These depend on the instrument and mode of observation.
In general, it should be mentioned if necessary, arcs and flats at night for spectroscopy, flux spectroscopic standards on the same night, and stars for telluric correction.
\end{enumerate}

\section{Considerations for Band 3 programs} \label{s:banda3}

At the Gemini Observatory, programs in Band 3 are used to complete the observation queue when there are no programs available in the highest priority bands, called Band 1 and 2, in which the highest-rated proposals in the ranking are located.
Consequently, successful programs in Band 3, that is, those with the greatest probability of being observed, are those that require below-average observation conditions.
The ``Plan for Band 3'' section of the proposal allows technical adjustments to increase the chances of observation if the proposal falls into priority Band 3.

If the observation conditions requested in section TD are already relaxed, that is, IQ85 and/or CC80, the observation proposal does not need modifications. It is understood that the proposed science goals are achievable under these conditions.
This case should be indicated with a phrase similar to: {\it ``This program is suitable for Band 3. No further modifications are required.''}.

In contrast, if the proposal requests average observation conditions (IQ70 and/or CC50), and none of the proposed objectives is achievable under relaxed conditions (IQ85 and/or CC80), then the program is not recommended for Band 3.
This should be indicated with a phrase similar to:
{\it ``This program is not suitable for Band 3''}.

Finally, if one or more objectives were achievable under relaxed observation conditions, this section should specify the new observation conditions, what objectives would be achievable under these conditions, and all the technical changes necessary to carry it out, including: total and minimum time, instrument configuration, S/N, etc.
The content of this section should be analogous to that of the TD section (Section~\ref{sec_dt}), and it is required to attach an ITC output with the new configuration and observation conditions.

\section{Recommendations} \label{s:reminders}

The following is a list of reminders to avoid penalties.

\begin{itemize}
\item Follow the ``DARP'' (Dual Anonymous Review Process) 
compliant anonymous writing guidelines outlined in the document
\href{https://noirlab.edu/science/observing-noirlab/proposals/anonymization-instructions.pdf}{``Anonymization Instructions for PIs''} \citep{darp_full}.
A shorter version in Spanish can be found on the \textit{``Oficina Gemini Argentina''} 
\href{https://www.argentina.gob.ar/ciencia/sact/convocatorias-gemini}{ \it call for proposals} website \citep{oga_web}.

\item Respect page limits and formatting.

\item Thoroughly check the instrument specifications and the observing strategies to prevent missing information in the TD.

\item Include the ITC output where the expected S/N for the sample objects is demonstrated, with the same configuration described in the TD.

\item If there is no ITC available for the requested instrument, it must be explained in as much detail as possible how the estimation of exposure times was made to obtain the desired S/N, since the Technical Office must be able to reproduce the calculation of the requested time with the information provided.

\item In joint proposals, the distribution of time must seek a balance between the number of collaborators per country/partner and the fraction of time requested by each country/partner.
It is recommended to align within one of the following two scenarios:
\begin{itemize}
 \item[a)] If the number of collaborators per country/partner is similar, it is recommended that the fraction of time requested be proportional to the available time of the corresponding country/partner.
 \item[b)] If the number of collaborators from one country/partner is significantly greater than that from another country/partner, it is recommended that the time requested be proportional to the number of collaborators from the corresponding country/partner.
\end{itemize}

\item Check if there are identical observations to those in the proposal (the same object, instrument, configuration, conditions, etc.) in the \href{https://archive.gemini.edu/searchform}{Gemini Observatory Archive} (GOA) database. In case there is no match, add the phrase {\it ``The GOA search did not reveal duplicate observations''}.

\item If there were previous observations in the GOA, it is essential to justify the need for new data (e.g. temporal sampling of the object, problems with previous data, etc.).

\item Consider including the most representative figures and graphs for the observation proposal. Figures offer the opportunity to present evidence of the scientific case (e.g. ``xy graphs'', histograms, previous results, etc.), of the method of analysis, and of the design and planning of the observations (image of the field, example of the sample, etc.).
They are also useful to explain the scientific contribution of the project and demonstrate the value of the observations.

\item Check the availability of guide stars for the selected observing conditions and sky area, especially if the program requires a specific position angle, for example for spectroscopy slit orientation.
The absence of guide stars could be a reason for cancellation of the program.

\item Carry out a critical reading of the proposal before being sent, verifying its completeness, neatness and writing.

\item Finally, it is recommended that the group of collaborators review the proposal according to the valuation system of the rubric in the following section.
\end{itemize}

\section{Rubric} \label{s:rubric}
The table below provides a valuation system
suggested for reviewing or planning the content of each section
in a Gemini Observatory telescope time proposal.
It is a tool to identify the current state of a presentation
and find possible aspects to improve.

\begin{table*}
\centering
 \makebox[0.5\textwidth][c]{}
 \makebox[0.5\textwidth][c]{{\Large Scientific Justification}}
 \makebox[0.5\textwidth][c]{}
 \begin{tabular}{p{0.15\linewidth} | p{0.8\linewidth}}
 \hline \\
 
 {\large \bf Rating} & {\large \bf \ref{context_cient}- Scientific Context} \\
 \hline
 High & The context is clear and complete. You can identify the general theme and the specific unknowns associated with the problem you are trying to address. The information is presented in order, from the general to the specific, and in a coherent manner, correctly using the relevant bibliographic citations, which helps to understand and assess the objectives. Reading is fluid and the case is understood with a quick read. It contains images and/or tables that help explaining the topic and/or problem.\\
 \hline
 Mid & The context contains minimal information and/or is difficult to read because it is not clear enough, disordered, or incomplete. Reading is not fluid. The general theme is identified but the problem that is being addressed is not. Bibliographic references are missing.
\\
 \hline
 Low & No problem is raised, the information is disordered or incomplete so there is no introduction that leads to the statement of objectives. It is a context made up of loose ideas. It is difficult to read. The figures and/or tables are not clear, and/or they do not help to understand the problem.
 \\
 \hline
 Absent & Does not contain scientifically relevant information. No specific topic is addressed. It does not contain figures or tables that help understand the science case. It does not contain bibliographic references.
 \\
 \hline \\
{\large \bf } & {\large \bf \ref{relevance}- Problem Relevance } \\
 \hline
 High & The topic addressed has strong connections with other areas of research and the results would contribute to other areas. The topic is trendy and has the support of a large part of the community. The proposed project is capable of contributing to, or answering, open questions in the area.
 \\
 \hline
Mid & Contribution is valuable for other areas, but is not essential.
The topic is not new but it seeks to improve something already known.
\\
 \hline
Low & The theme is very specific. Contribution is limited to minor details.
The project is a pilot test.
\\
\hline
 Absent & The proposal is to reproduce widely known results from previous studies.
 \\
 \hline \\
 {\large \bf } & {\large \bf\ \ref{obj_generla}- General Objective } \\
 \hline
 High & The general objective is clear and well connected to the context provided. The contribution of the general objective to the topic is significant.\\
 \hline
 Mid & The general objective is difficult to understand. It appears disconnected from the context. A limited contribution is indicated.
 \\
 \hline
 Low & The general objective is not clear and/or is distributed throughout the text. It is presented repeatedly, but without a logical connection with the rest of the text or without delving into its justification.
 \\
 \hline
 Absent & No general objective is set.
 \\
 \hline \\ 
 {\large \bf } & {\large \bf\ \ref{obj_espec}- Specific or Immediate Objectives } \\
 \hline
 High & The objectives are presented explicitly and are broken down from the general objective. They are seen as intermediate steps towards the general objective. They are feasible (achievable) objectives. They are specific individual contributions.
 \\
 \hline
 Mid & The objectives presented are incomplete. Some seem connected and moving towards the general objective, but others do not. There is a very strong dependency between one and the other; if one objective were to fail, it would not be possible to advance in others.
 \\
 \hline
 Low & Specific objectives are not clear or are confusingly distributed throughout the text. They are not connected to the overall goal.
 \\
 \hline
 Absent & No specific objectives are set.
 \\
 \hline \\

{\large \bf }& {\large \bf \ref{impact}- Impact } \\
 \hline
 High & The results of the proposed research could change and/or revolutionize the ``current vision'' that we have on the topic raised. The scope of the results affects various areas of science.
These are record-breaking objects (more massive, more distant, new discoveries, etc.).
A scientifically contradictory and/or controversial topic is addressed in an intelligent and convincing manner. Solid evidence is sought for a revolutionary theme. The results could cause big changes in perspective in the community. Extraordinary evidence of a little-known but key phenomenon is sought to consolidate or discard a theory.
\\
 \hline
 Mid & It is an interesting topic but it will not affect the vision of the topic raised.
 The results will only reach a specific area of astronomy.
 The phenomenon is not well understood, and while there is limited debate about its nature, the contribution is interesting but not critical or fundamental.
 \\
 \hline
 Low & There are no controversies or extraordinary contributions to the topic. This is not a little-known phenomenon. These are not record-breaking objects.
\\
 \hline
 Absent & The topic is widely studied and understood.
 \\
 \hline \hline
 \end{tabular}
\label{t:JC2}
\end{table*}

\begin{table*}
\centering
 \makebox[0.5\textwidth][c]{}
 \makebox[0.5\textwidth][c]{{\Large Experimental Design}}
 \makebox[0.5\textwidth][c]{}
 \begin{tabular}{p{0.15\linewidth} | p{0.8\linewidth}}
 \hline \\
{\large \bf Rating} & {\large \bf \ref{sample}- Sample } \\
 \hline
 High & The sample justification provided is aligned with the science theme.
 The experiment could not be carried out with other objects. The sample selection is solid.\\
 \hline
 Mid & The sample selection is arbitrary but contributes to the scientific case. This is a test of a selection criterion. The sample is logical, but the selection is not clear.\\
 \hline
 Low & No reasons or criteria are provided for sample selection.
 \\
 \hline
 Absent & The sample is not described.
 \\
 \hline \\
{\large \bf }& {\large \bf\ \ref{strategy}- Strategy} \\
 \hline
 High & A complete description of the strategy and data management is provided, which is aligned with the scientific objectives. The strategy is solid and has been tested in previous studies. The explanation is clear and well-founded with bibliographical citations. \\
 \hline
 Mid & The strategy is known and correct for the scientific case but no justifications or detailed explanations of the process are provided.\\
 \hline
 Low & The strategy is reasonable but not entirely clear. It does not align with the objectives. It has errors. \\
 \hline
 Absent & No details of the proposed work and measurements are presented.\\
 \hline \\
{\large \bf }& {\large \bf\ \ref{tools}- Analysis tools }\\
 \hline
 High & The tools for data analysis are correctly described, including specific software, comparison with theoretical models, etc.\\
 \hline
 Mid & The tools involved in the analysis are mentioned, but no details are given. \\
 \hline
 Low & The analysis tools are too rudimentary for the proposed measurements, which would result in insufficient precision. \\
 \hline
 Absent & Data analysis tools are not proposed. \\
 \hline \\
{\large \bf }& {\large \bf\ \ref{contribution}- Contribution to the General Objective } \\
 \hline
 High & There is a clear relation between the strategy, the proposed measurements, and the general objective. It is demonstrated how the proposed measurements would contribute to the resolution of the problem raised in the Scientific Justification. The strategy presentation returns to the objectives to demonstrate the contribution to the topic. \\
 \hline
 Mid & Only part of the measurements and strategy are linked to the objectives. \\
 \hline
 Low & The connection between the strategy and the objective is not appreciated.
 \\
 \hline
 Absent & The strategy does not correspond to the stated objectives.
\\
 \hline \hline
\end{tabular}
\label{t:DE}
\end{table*}

\begin{table*}
\centering
 \makebox[0.5\textwidth][c]{}
 \makebox[0.5\textwidth][c]{{\Large Technical Description}}
 \makebox[0.5\textwidth][c]{}
 \begin{tabular}{p{0.15\linewidth} | p{0.8\linewidth}}
 \hline \\
{\large \bf Rating}& {\large \bf\ \ref{configuration}- Instrument configuration } \\
 \hline
 High & The description of the instrument, its configuration, and the use of a 8.1 m diameter telescope are duly justified and complete. All the technical aspects that must be taken into account to carry out the observation are carefully detailed.
 The configuration matches the ITC output presented.
 \\
 \hline
 Mid & The main configurations of the instrument to be used are described, but important details are omitted. The configuration matches the ITC output presented.
 \\
 \hline
 Low & Not all the information necessary for the complete configuration of the instrument is provided and/or some fundamental characteristic is omitted, and/or the configuration does not correspond with the ITC outputs presented, or the needs raised in the ED section.\\
 \hline
 Absent & Most instrument configuration is omitted and ITC outputs are not attached or do not correspond to program needs.
\\
 \hline \\
{\large \bf } & {\large \bf \ref{resolution}- Angular and/or spectral resolution } \\
 \hline
 Justified & The angular and/or spectral resolution achievable with the configuration and observing conditions requested is mentioned, and it is compared with the limits necessary for the scientific objective. \\
 \hline
 Unjustified & It is not demonstrated that the achievable resolution will be sufficient. \\
 \hline
 Absent & Angular and/or spectral resolution is not mentioned.
 \\
 \hline \\
 & {\large \bf \ref{condition}- Observing Conditions } \\
 \hline
 Justified & The selected observing conditions are explicitly justified to meet the ED requirements.
 Cloudiness (CC) is justified by the need for flux calibration.
 Image quality (IQ) with the need for spatial resolution.
 The sky brightness (SB) and air mass with the magnitude of the source and the spectral range of interest.
 The best conditions are justified to achieve reasonable observation times.
 
 \\
 \hline
 Unjustified & The conditions used in the ITC are mentioned, but no choice is justified.
 \\
 \hline
 Absent & Observing conditions are not mentioned.
 \\
 \hline \\
{\large \bf } & {\large \bf \ref{offsets}- \emph{Offsets}, \emph{nodding} or \emph{dithering}} \\
 \hline
 Justified & The description of the pattern of \textit{offsets} in the sky is clear and reasonable for the type of observation.
 The need (or not) for \textit{nodding} along the slit (e.g. ABBA pattern), or to a region of the sky free of bright sources when observing extended sources, is justified. The \textit{offsets} in central wavelength and the \textit{offsets} for the assembly of the final mosaics of the image are justified.\\
 \hline
 Unjustified & The types of necessary \textit{offsets/nodding/dithering} are mentioned but not justified. Or, the type of \textit{offset/nodding/dithering} proposed is not compatible with the ED and calls into question the scientific objectives.\\
 \hline
 Absent & No indication is given as to what the appropriate pattern of \textit{offsets/nodding/dithering} would be for the type of observation.\\ 
 \hline \\
{\large \bf } &
{\large \bf\ \ref{s/n}- Signal to Noise Ratio} \\
 \hline
 High& The S/N was calculated with the ITC and coincides with the S/N necessary for the measurements described in the ED.\\
 \hline
 Mid & The necessary S/N is indicated, but does not match that in the examples calculated with the ITC.\\
 \hline
 Low & Very high S/N values are requested without giving a justification. The proposed S/N does not correspond to the examples calculated with the ITC.
 \\
 \hline
Absent & The S/N necessary to achieve the objectives is not mentioned.
 \\
 \hline \\

{\large \bf } & {\large \bf \ref{times}- Total and minimum times} \\
 \hline
 High & The total and minimum times have been calculated correctly. The necessary \textit{overheads} and time for calibrations have been included (except when the \textit{baseline} calibrations provided by the observatory were sufficient).
 The times coincide with the ITC examples or were calculated with a simulation of the observations in the Observing Tool.
The reason for the total time and the minimum time is justified (for example, to reach certain S/N, to observe a given number of objects, etc.).\\
 \hline
 Mid & The times provided as total and minimum are incorrect but have been updated
 after the Gemini Office technical report.\\
 \hline
 Low & No arguments have been given to justify the total time. It is not indicated what is expected to be achieved with the minimum time.
 \\
 \hline
 Absent & Total and minimum times are not indicated.\\
 \hline \\
{\large \bf } & {\large \bf \ref{calibrations}- Calibrations} \\
 \hline
 High & All calibrations relevant to the observation type are requested. Standard stars are requested for flux calibration when the ED requires it. The number of stars necessary for telluric correction in infrared spectroscopy is indicated.\\
 \hline
 Mid & Some necessary calibrations have been omitted (e.g. no flux standards included but the scientific case depends on it), but have been updated
 after the Gemini Office technical report.
\\
 \hline
 Low & The requested calibrations are insufficient.\\
 \hline
 Absent & No mention of required calibrations, although the scientific case depends on them. \\
 \hline \hline
\end{tabular}
\label{t:DT2}
\end{table*}

\section*{Acknowledgements}

We wish to recognize the valuable contribution
of the User Advisory Committee
from Argentina, Dr. Andrea Ahumada, Dr. Carlos Saffe,
Dr. Guillermo Hägele, and
Dr. Ana Pichel, for all the comments
received in the preparation of this manuscript.
 We also wish to express our total gratitude to the Gemini Argentina Office,
Dr. Luciano Garcia, Dr. Gabriel Ferrero,
Dr. Carlos Escudero, and Dr. Leandro Sesto,
for all the productive discussions that we share,
and for the suggestions that have greatly enriched this document.
This article uses the Latex template NobArticle by
José António Portela Areia.

\printbibliography


\end{document}